# Fundamental limit to the rectification of near-field heat flow: The potential of intrinsic semiconductor films

Qizhang Li[1,2], Qun Chen[1†] and Bai Song[2,3,4†]

[1]*Key Laboratory for Thermal Science and Power Engineering of Ministry of Education, Department of Engineering Mechanics, Tsinghua University, Beijing 100084, China*

[2]*Beijing Innovation Center for Engineering Science and Advanced Technology, Peking University, Beijing 100871, China*

[3]*Department of Energy and Resources Engineering, Peking University, Beijing 100871, China*

[4]*Department of Advanced Manufacturing and Robotics, Peking University, Beijing 100871, China*

[†]Corresponding author. Email: chenqun@tsinghua.edu.cn, songbai@pku.edu.cn

**ABSTRACT**

We derive the fundamental limit to near-field radiative thermal rectification mediated by an intrinsic semiconductor film within the framework of fluctuational electrodynamics. By leveraging the electromagnetic local density of states, we identify $\varepsilon''_\text{H}/\varepsilon''_\text{L}$ as an upper bound on the rectification magnitude, where $\varepsilon''_\text{H}$ and $\varepsilon''_\text{L}$ are respectively the imaginary parts of the film permittivity at high and low temperatures. This bound is tight and can be approached regardless of whether the film is suspended or supported. For intrinsic silicon the limit can in principle exceed $10^9$. Our work highlights the possibility of controlling heat flow as effectively as electric current, and offers guidelines to potentially achieve this goal.

Nonreciprocal transport is not only of great theoretical interest but also serves as the cornerstone of modern electronics industry [1]. In analogy to electrical diode, a thermal diode rectifies the flow of heat [2–5] and enables a range of emerging applications in thermal management, energy harvesting, and thermal circuits [6,7]. The first observation of thermal rectification is usually attributed to Starr's experiment in 1936 on heat conduction across a copper-cuprous oxide interface [8]. Since then, thermal diodes mediated by various energy carriers such as phonons [2–4,9–12], electrons [13–15], and photons [5,16–21] have been extensively explored, with a collective focus on novel mechanisms and materials to achieve ever higher rectification performance. To date, however, the magnitude of thermal rectification remains substantially lower than what is routinely available with a typical electrical diode. More importantly, the fundamental limit to the rectification magnitude remains unexplored due to an overwhelmingly large parameter space, with dimensions including the mechanisms, materials, geometries, and working temperatures.

Here, we make an effort to address this longstanding challenge. We focus on thermal diodes mediated by photon tunneling in the near field which can potentially achieve ultrahigh rectification over a broad temperature range [17,22–29]. During the past decade, near-field radiative thermal diodes have been widely studied, with the necessary nonlinearity [30] provided in particular by semiconductors [22,29,31–35] and phase transition materials [25–28,36,37] via their temperature-dependent permittivity. Among various geometries, the parallel-plane configuration is of central importance considering its fabricability and scalability, in addition to its ability to substantially rectify a large heat flow [38–41]. State-of-the-art parallel-plane diodes designed with phase transition materials theoretically offer over 2-orders-of-magnitude rectification [26–28]. Very recently, by using a thin film of intrinsic silicon (*i*-Si) in the parallel-plane geometry, ~5-orders-of-magnitude rectification has been predicted [29], which represents more than 100-fold improvement. This encouraging progress motivated us to wonder what the limit might be.

In what follows, we derive the fundamental limit to near-field thermal rectification enabled by an intrinsic semiconductor film, using the electromagnetic local density of states (LDOS) [42] as an intermediary. Briefly, we consider a diode (Fig. 1a) with one terminal incorporating the semiconductor film to provide a substantial LDOS contrast as the temperature varies, while the other serving as a temperature-insensitive bandpass filter that limits the heat flow to a desired spectral range [26]. We first show that, when comparing the film at high and low temperatures, the contrast of near-field LDOS is bounded by that of the imaginary part of the permittivity ($\varepsilon_H''/\varepsilon_L''$). Subsequently, we further demonstrate that $\varepsilon_H''/\varepsilon_L''$ is a tight limit to the rectification magnitude regardless of whether the film is freely suspended or supported on a substrate. Finally, using *i*-Si as an example, we reveal that thermal-rectification magnitude over $10^9$ could in principle be achieved, which matches or even exceeds that of electrical diodes [43].

In our thermal diode, the nonlinearity originates from the temperature-dependent complex permittivity ($\varepsilon$) of the intrinsic semiconductor. As temperature rises, more free carriers are thermally excited which give rise to a substantial increase in the imaginary part of the permittivity ($\varepsilon''$), especially in the infrared. Meanwhile, the real part ($\varepsilon'$) is dominated by the lattice and is usually much less affected (Fig. 1b) [44,45]. This general feature allows us to consider a representative permittivity in the form of $\varepsilon(T) = \varepsilon' + i\varepsilon''(T)$ for our theoretical analysis.

The basic structure of our radiative thermal diode (Fig. 1a) is proposed based on a LDOS-centered rational design approach [26,27,29], which enables us to separately analyze each of the two terminals and is key to our derivation of the upper limit to thermal rectification. With such a design, the rectification magnitude closely follows the LDOS contrast provided by the semiconductor film [26,29]. As a result, we begin by considering the LDOS $\rho(\omega, z)$ at a small distance $z$ above the film [42]. Specifically, we focus on the wavevector-resolved local density of states (WLDOS), $\rho_w(\omega, \kappa, z)$, defined by $\rho(\omega, z) = \int \rho_w(\omega, \kappa, z) d\kappa$ where $\kappa$ is the parallel wavevector [46,47]. For the $p$-polarized evanescent modes which usually dominate the near-field heat flow [46,48], the WLDOS is given by [49,50]

$$\rho_w(\omega, \kappa, z) = \frac{\kappa^3}{2\pi^2 \omega |\gamma_0|} e^{-2\mathrm{Im}(\gamma_0)z} \mathrm{Im}(R), \tag{1}$$

where $\gamma_0 = \sqrt{k_0^2 - \kappa^2}$ is the normal wavevector in vacuum and $k_0 = \omega/c$; $R$ is the reflection coefficient of the film and $\mathrm{Im}(R)$ is the generalized emissivity [51]. Further, since the modes propagating inside the film ($\kappa < \mathrm{Re}(\sqrt{\varepsilon})k_0$) have been largely suppressed by the very limited source volume [26,29,52], only the high-$\kappa$ modes ($\kappa \gg k_0$) are of importance.

According to Eq. (1), the contrast of WLDOS ($\rho_{w,H}/\rho_{w,L}$) between high ($T_H$) and low ($T_L$) temperatures is given by the ratio of the corresponding film emissivities, i.e.,

$$\frac{\rho_{w,H}}{\rho_{w,L}} = \frac{\mathrm{Im}(R_H)}{\mathrm{Im}(R_L)}. \tag{2}$$

For a suspended film with $\varepsilon_H'' > \varepsilon_L''$ and $\varepsilon' > 1$, we find that the emissivity contrast $\mathrm{Im}(R_H)/\mathrm{Im}(R_L)$ for the high-$\kappa$ modes is consistently smaller than $\varepsilon_H''/\varepsilon_L''$ (see Supplementary Material for details [53]), which further leads to

$$\frac{\rho_{w,H}}{\rho_{w,L}} < \frac{\varepsilon_H''}{\varepsilon_L''}. \tag{3}$$

Since Eq. (3) holds for every high-$\kappa$ mode, the LDOS contrast ($\rho_H/\rho_L$) upon integration of the WLDOS is also bounded by $\varepsilon_H''/\varepsilon_L''$, regardless of the distance and film thickness. For intrinsic semiconductors which usually feature a much smaller $\varepsilon''$ in the infrared compared to $\varepsilon'$ (see Fig. S1 for $i$-Si as an example) [44,53–55], we further find that $\rho_H/\rho_L$ can be well approximated by $\varepsilon_H''/\varepsilon_L''$ [53]. In Fig. 1b, we plot exact calculations of the LDOS contrast above a suspended film with a hypothetical permittivity, which agree well with the theoretical analysis. In addition, we show results for a suspended $i$-Si film at three representative frequencies with increasing $T_H$ (inset of Fig. 1b), which again validates our derivation.

Now that a limit to the LDOS contrast is established, we proceed to analyzing the rectification potential of the proposed thermal diode by bringing a bandpass filter to the near field of the suspended semiconductor film. Based on fluctuational electrodynamics [56], the spectral heat flux at frequency $\omega$ between two planes across a vacuum gap $d$ can be written as [48,57]

$$q(\omega) = \left[\Theta(\omega,T_H) - \Theta(\omega,T_L)\right] \sum_{j=s,p} \int_0^\infty \frac{\kappa d\kappa}{4\pi^2} \tau_j(\omega,\kappa), \tag{4}$$

where $\Theta(\omega,T)$ is the mean energy of a harmonic oscillator, $j$ represents the polarization, and $\tau$ is the transmission probability which for an evanescent mode is given by

$$\tau(\omega,\kappa) = \frac{4\operatorname{Im}(R_f)\operatorname{Im}(R)e^{-2\operatorname{Im}(\gamma_0)d}}{\left|1 - R_f R \exp(2i\gamma_0 d)\right|^2}. \tag{5}$$

Here, $R_f$ is the reflection coefficient of the bandpass filter and is temperature independent. Comparing the forward and reverse biased scenarios, the contrast of the transmission probability is thus

$$\frac{\tau_F}{\tau_R} = \frac{\text{Im}(R_H)}{\text{Im}(R_L)} \frac{\left|1 - R_f R_L \exp(2i\gamma_0 d)\right|^2}{\left|1 - R_f R_H \exp(2i\gamma_0 d)\right|^2}. \tag{6}$$

One can see that $\tau_F/\tau_R$ is related to $\text{Im}(R_H)/\text{Im}(R_L)$ by a complicated term that depends on the bandpass filter, and therefore may not be solely determined by the semiconductor film. However, for intrinsic semiconductors, we find that $|1 - R_f R_H \exp(2i\gamma_0 d)| > |1 - R_f R_L \exp(2i\gamma_0 d)|$ always holds for $p$-polarized high-$\kappa$ modes, again by considering that $\varepsilon''$ is much smaller than $\varepsilon'$ [53]. This, combined with Eqs. (2), (3), and (6), leads to

$$\frac{\tau_F}{\tau_R} < \frac{\rho_{w,H}}{\rho_{w,L}} < \frac{\varepsilon''_H}{\varepsilon''_L}, \tag{7}$$

regardless of $R_f$. With Eq. (4), the contrast of the spectral heat flux should satisfy

$$\frac{q_F}{q_R} < \frac{\varepsilon''_H}{\varepsilon''_L}. \tag{8}$$

Finally, we arrive at the following relation for the contrast between the total heat fluxes:

$$\frac{Q_F}{Q_R} = \frac{\int_0^\infty q_F d\omega}{\int_0^\infty q_R d\omega} < \frac{\int_0^\infty \left(\frac{\varepsilon''_H}{\varepsilon''_L}\right)_{\max} q_R d\omega}{\int_0^\infty q_R d\omega} = \left(\frac{\varepsilon''_H}{\varepsilon''_L}\right)_{\max}. \tag{9}$$

According to Eq. (9), the rectification magnitude of the thermal diode is ultimately limited by the maximum of $\varepsilon''_H/\varepsilon''_L$ that is accessible in the frequency range of interest for the intrinsic semiconductor.

To validate the theoretical results above, in particular the spectral relation Eq. (8), we perform exact calculations for a diode pairing the semiconductor film with a narrowband filter [26,29] which essentially limits the heat flow to the vicinity of a particular frequency. Here, we employ a polar dielectric filter which features a sharp LDOS peak due to its surface phonon polaritons (SPhPs) in the so-called Reststrahlen band [26]. The permittivity of polar dielectrics is often described by the Lorentz model as $\varepsilon_\mathrm{d} = \varepsilon_\infty (\omega^2 - \omega_\mathrm{LO}^2 + i\Gamma\omega)/(\omega^2 - \omega_\mathrm{TO}^2 + i\Gamma\omega)$ [58], and can be considered temperature independent when compared to intrinsic semiconductors [32,59]. The LDOS peak of the polar dielectric occurs at the resonant frequency $\omega_\mathrm{r}$ where $\varepsilon_\mathrm{d}' = -1$ [42], which translates the corresponding LDOS contrast provided by the semiconductor film into the rectification performance of the diode (Fig. 2a) [29]. In order to minimize contributions from outside the Reststrahlen band, we use a thin film instead of bulk polar dielectric to suppress the propagating modes inside the medium [29]. And the film thickness is set to be larger than the penetration depth of the high-$\kappa$ modes (comparable to $z$) to avoid splitting of the LDOS peak which occurs when the SPhPs couple within the film [58,60,61].

Without loss of generality, we consider three hypothetical polar dielectrics with $\omega_\mathrm{r} = 1 \times 10^{13}, 5 \times 10^{13}$, and $1 \times 10^{14}$ rad s$^{-1}$ to pair with the $i$-Si film, which are representative of real materials such as potassium bromide, barium fluoride, and magnesium oxide [55]. As shown in Fig. 2a, the contrasts of the spectral heat fluxes for all three diodes are clearly limited by $\varepsilon_\mathrm{H}''/\varepsilon_\mathrm{L}''$ of $i$-Si, as predicted by Eq. (8). Remarkably, the three $q_\mathrm{F}/q_\mathrm{R}$ curves almost completely overlap with each other, and closely approach $\varepsilon_\mathrm{H}''/\varepsilon_\mathrm{L}''$ over the entire frequency range. This suggests that $\varepsilon_\mathrm{H}''/\varepsilon_\mathrm{L}''$ may be a tight bound to $q_\mathrm{F}/q_\mathrm{R}$, which we theoretically confirm in the Supplementary Material [53]. When it comes to the total heat fluxes $Q_\mathrm{F}$ and $Q_\mathrm{R}$, the spectral fluxes at around $\omega_\mathrm{r}$ contribute dominantly (>95% from within the Reststrahlen bands). As a result, the contrast $Q_\mathrm{F}/Q_\mathrm{R}$ is roughly given by $q_\mathrm{F}/q_\mathrm{R}$ at $\omega_\mathrm{r}$ and can also be well approximated by the corresponding $\varepsilon_\mathrm{H}''/\varepsilon_\mathrm{L}''$ (Fig. 2b). This concludes our derivation and validation of the limit to thermal rectification mediated by a suspended intrinsic semiconductor film.

We now move on to analyze the case of an intrinsic semiconductor film supported on a bulk substrate. For this configuration, we find that the WLDOS contrast also satisfies

$$\left(\frac{\rho_{w,H}}{\rho_{w,L}}\right)_s = \left[\frac{\text{Im}(R_H)}{\text{Im}(R_L)}\right]_s < \frac{\varepsilon_H''}{\varepsilon_L''}, \tag{10}$$

regardless of the substrate permittivity $\varepsilon_s$ [53]. Therefore, the LDOS contrast is again limited by $\varepsilon_H''/\varepsilon_L''$, same as in the case of a suspended film. This is confirmed by calculations of the LDOS contrast for a hypothetical semiconductor film ($\varepsilon_H''/\varepsilon_L'' = 10^4$) sitting on a substrate the permittivity of which is systematically varied, as shown in Fig.3a. Notably, for substrate with $\varepsilon_s' \in (-\varepsilon', -1)$, the LDOS contrast is consistently very small. This is because resonant surface modes are excited at the film-substrate interface which overwhelm the original LDOS characteristics of the film [53]. Therefore, in order to achieve large thermal rectification, substrates with such permittivity should be avoided (Fig. S2). For a substrate with a proper $\varepsilon_s'$ and a sufficiently small $\varepsilon_s''$, the LDOS contrast gets very close to $\varepsilon_H''/\varepsilon_L''$. Accordingly, we further find that the limits to thermal rectification via a suspended film [Eqs. (7)-(9)] all apply to the case of a substrate-supported film [53]. In particular, a maximum rectification magnitude very close to the limit of $\varepsilon_H''/\varepsilon_L''$ can also be achieved when the semiconductor film is supported by a transparent substrate ($\varepsilon_s'' \approx 0$) [26], as revealed in both theoretical analysis [53] and numerical calculations (Fig. 3b).

With $\varepsilon_H''/\varepsilon_L''$ established as a tight limit to the rectification magnitude of a thermal diode mediated by an intrinsic semiconductor film, we now explore the potential of *i*-Si by examining its permittivity as a function of temperature and frequency (Fig. 4). By setting $T_L$ to 300 K and increasing $T_H$ up to 1000 K, we observe a large parameter space characterized by over 3-orders-of-magnitude permittivity contrast. The maximum rectification attainable by a diode though, depends on the bandpass filter that determines the spectral range dominating thermal transport.

With a narrowband filter, one can potentially read the $\varepsilon_H''/\varepsilon_L''$ map as a map of maximum rectification. For example, by using a polar dielectric film (potassium bromide etc.), we obtain rectification magnitude very close to the maximum $\varepsilon_H''/\varepsilon_L''$ within its Reststrahlen band where SPhPs can be excited, as marked by the dashed horizontal lines in the inset of Fig. 4. The case of broadband filter is quantitatively less straightforward but conceptually the same.

Intriguingly, with $T_L = 300$ K and $T_H = 1000$ K, the maximum $\varepsilon_H''/\varepsilon_L''$ of $i$-Si reaches extraordinarily large values of over $10^9$ at frequencies around the bandgap $\omega_g$ (Fig. 4), which remains larger than $10^4$ even when $T_H$ is lowered to 500 K. This is because $i$-Si features very weak lattice absorption around $\omega_g$, so that the variation of carrier concentration can induce a dramatic change of $\varepsilon''$ (Fig. S1) [29,45]. To translate the extraordinary permittivity contrast into an unprecedented thermal rectification though, requires a matching narrowband filter in the near-infrared, which is beyond the scope of the present study. At the opposite side of the spectrum, $\varepsilon_H''/\varepsilon_L''$ is also relatively large (~$10^5$) but of a more broadband nature, which can be readily utilized by using a broadband filter such as doped-silicon ($d$-Si), as demonstrated in Ref. [29]. The case of $d$-Si further highlights the potential of temperature-dependent filters, which allows rectification magnitude greater than the contrast provided solely by the intrinsic semiconductor. One promising possibility is to use tailored metal-to-insulator transition materials [37].

In summary, by employing a LDOS-based approach, we have shown that $\varepsilon_H''/\varepsilon_L''$ is a tight limit to near-field thermal rectification mediated by an intrinsic semiconductor film, whether it is suspended or supported on a substrate. This allows one to conveniently but reliably evaluate the rectification potential of various materials without actually modeling the thermal transport process. We show that over 9-orders-of-magnitude thermal rectification is in principle possible with $i$-Si.

Our work reveals an opportunity to achieve unprecedentedly high thermal rectification comparable to that of electrical diodes. Before we conclude, it is tempting to ask why $\varepsilon_\text{H}''/\varepsilon_\text{L}''$ is the limit? Intuitively, the answer may lie in the fluctuation-dissipation theorem [56,57,62] which demands the current correlation that determines electromagnetic emissions to be simply proportional to the imaginary part of the material permittivity.

**ACKNOWLEDGEMENT**

This work is supported by the National Natural Science Foundation of China (Grants Nos. 52076002 and 52125604), the Beijing Innovation Center for Engineering Science and Advanced Technology, the XPLORER PRIZE from the Tencent Foundation, and the High-performance Computing Platform of Peking University.

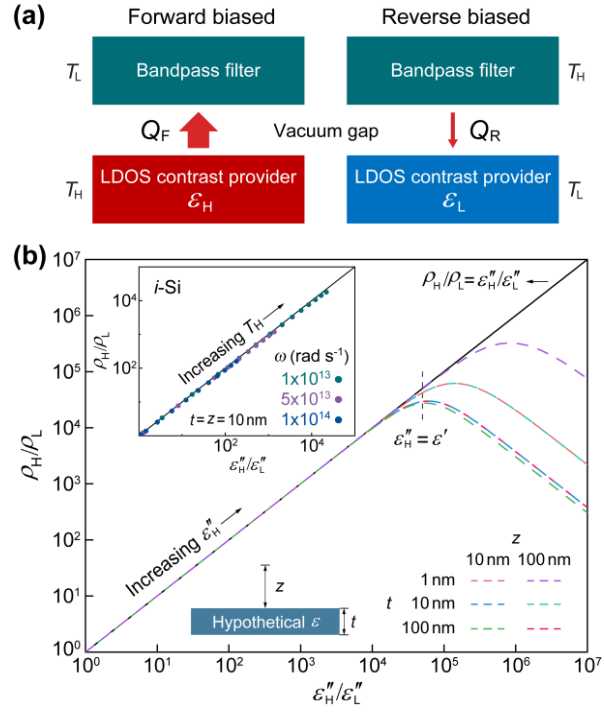

**Fig. 1. Conceptual design and LDOS analysis.** (a) Schematic of the thermal diode. (b) LDOS contrast of *p*-polarized evanescent modes as a function of $\varepsilon_H''/\varepsilon_L''$ at $\omega = 1 \times 10^{13}$ rad s$^{-1}$ for a hypothetical thin film. Here, $\varepsilon' = 10$ and $\varepsilon_L'' = 2 \times 10^{-4}$ (close to that of *i*-Si at 300 K) are fixed while $\varepsilon_H''$ is varied. Inset shows the case of an *i*-Si film at three representative frequencies, incorporating all contributing modes.

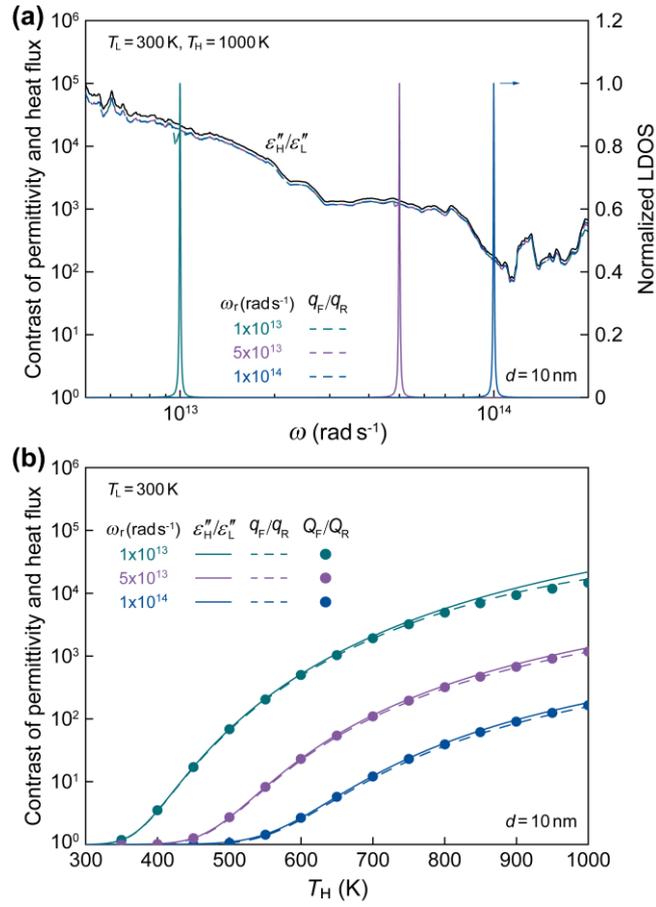

**Fig. 2. Thermal diode pairing a suspended *i*-Si film with a polar dielectric film.** (a) Contrast of $\varepsilon''$ and $q$ as a function of $\omega$, and normalized LDOS above the polar dielectric films. (b) Contrast of $\varepsilon''$ and $q$ at $\omega = \omega_r$, and contrast of $Q$, as a function of $T_H$. The *i*-Si film is 10 nm thick, and the polar-dielectric film is 100 nm thick. For the hypothetical polar dielectrics, we assume $\varepsilon_\infty = 1$, $\omega_{LO} = 1.2\omega_{TO}$, and $\Gamma = 0.01\omega_{TO}$.

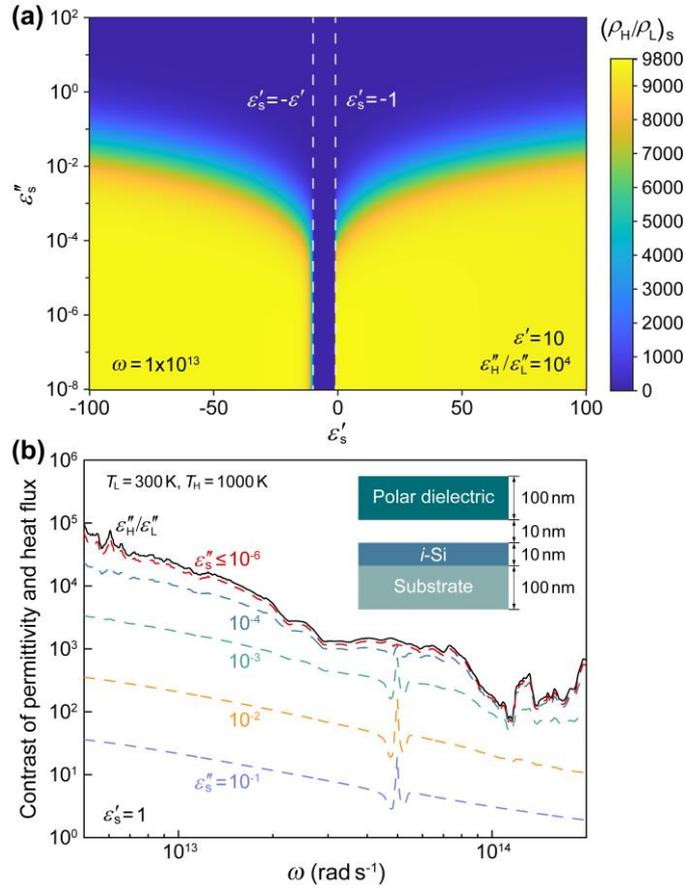

**Fig. 3. Thermal diode with the semiconductor film supported on a substrate.** (a) LDOS contrast map at 10 nm above a 10-nm-thick semiconductor film on a bulk substrate, as a function of $\varepsilon_s'$ and $\varepsilon_s''$. (b) Contrast of $q$ when pairing a polar dielectric film ($\omega_r = 5 \times 10^{13}$ rad s$^{-1}$) with a $i$-Si film on a thin substrate with fixed $\varepsilon_s'$ and varying $\varepsilon_s''$, as a function of $\omega$.

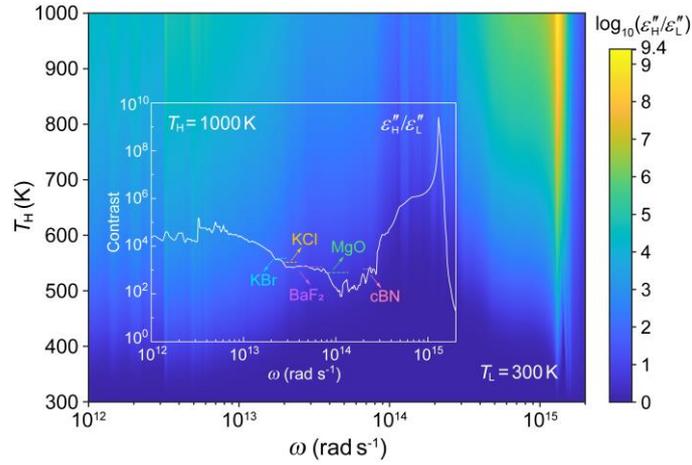

**Fig. 4. Thermal rectification potential of intrinsic silicon.** Contrast of $\varepsilon''$ for $i$-Si at different $\omega$ and $T_H$, with $T_L$ fixed at 300 K. Inset shows the case when $T_H = 1000$ K. Short dashed lines mark the optimized rectification magnitudes when pairing the $i$-Si film with a film of some representative polar dielectrics—potassium bromide, potassium chloride, barium fluoride, magnesium oxide, and cubic boron nitride [55]. The position and width of each dashed line mark the corresponding Reststrahlen band.